\documentclass[letter]{aa}  

\usepackage{graphicx}
\usepackage{txfonts}
\usepackage{lipsum}
\usepackage{subcaption}        
\usepackage{lscape}             
\usepackage{placeins}           
\usepackage[colorlinks,linkcolor=blue,citecolor=blue]{hyperref}
                                
\begin{document}

\title{Statistical evidence for linear combination frequencies in $\gamma$~Doradus stars being nonlinear resonant modes}

\titlerunning{Linear combination frequencies in $\gamma$~Dor stars}

 \author{Yanqi Mo\inst{1,2,3}
      \and
      Weikai Zong\inst{1,2,3}\fnmsep\thanks{Corresponding author: weikai.zong@bnu.edu.cn} 
      \and
       Xuan Wang\inst{4}
      \and
         Jian-Ning Fu\inst{1,2}
      \and
        Xiao-Yu Ma\inst{5}
        \and
        Mingfeng Qin\inst{6}
              \and
       St\'ephane Charpinet\inst{4}
   }

\institute{
    Institute for Frontiers in Astronomy and Astrophysics, Beijing Normal University, Beijing 102206, China
    \and
    School of Physics and Astronomy, Beijing Normal University, Beijing 100875, PR China
    \and
    International Centre of Supernovae (ICESUN), Yunnan Key Laboratory, Kunming, 650216, P. R. China
    \and
    IRAP, CNRS/Université de Toulouse/CNES, 14 Avenue Edouard Belin, Toulouse, 31400, France
    \and
     SpaceSciences, Technologies and Astrophysics Research (STAR) Institute, Université de Liège, Allée du 6 Août 19C, 4000 Liège, Belgium
     \and
     School of Physics, Nankai University, Tianjin 300071, China
   }

   \date{}

 \abstract
{The quest for the origin of linear combination frequencies in pulsating stars is critical for mode identification and subsequent seismic modeling. Hybrid $\gamma$~Doradus (Dor) stars provide a crucial testbed for such investigations due to their rich frequency content in both gravity (g) and pressure (p) modes. Here, we performed a comprehensive survey of 608 Kepler $\gamma$~Dor stars, detecting 81,265 frequencies, and identifying 1,895 harmonics and 45,968 combinations involving two parent frequencies. Our results suggest that the proportion of linear combinations increases with the frequency number in each $\gamma$ Dor star, leveling off at approximately 75\%. The majority of the parent-to-child amplitude ratios are distributed between 100 and 10,000, peaking between 2,000 and 4,000. Moreover, the amplitude ratios, involving two g-mode parents, are found to escalate with increasing combination frequency. These values are at least an order of magnitude higher than those predicted by nonlinearity distortion of the light curve, whereas they align well with those of nonlinear resonant coupling. Our findings suggest that a significant portion of combination frequencies are likely intrinsic modes under resonant conditions rather than products of surface geometric distortions, offering new insights that warrant systematic investigations across other types of pulsating star.
}


\maketitle
\nolinenumbers

\section{Introduction}
As a unique technique for probing interiors of pulsating stars, asteroseismology provides solid constraints on fundamental parameters and internal physics, such as rotation, mixing, and angular momentum transport \citep{2021RvMP...93a5001A,2022ARA&A..60...31K}. Driven by the era of big data, asteroseismic studies across various classes of pulsating stars have prompted many advances in reshaping modern astronomy \citep{2024A&A...692R...1A}. 
Crucially, pulsators can be probed down to their core regions if they present gravity (g) modes, making these oscillations vital for understanding deep stellar structures \citep{2011Natur.471..608B}.
Among these, $\gamma$~Doradus ($\gamma$~Dor) stars are intermediate-mass main-sequence A–F variables exhibiting high-order low-degree non-radial gravity (g) modes with periods ranging from several hours to a few days \citep{1999PASP..111..840K,1999MNRAS.309L..19H}. Their pulsations are generally attributed to the convective flux-blocking mechanism acting at the base of the convective envelope, while the instability strip is successfully reproduced by non-adiabatic models including time-dependent convection \citep{2000ApJ...542L..57G,2004A&A...414L..17D}.

$\gamma$~Dor stars are invaluable targets for asteroseismic investigations owing to their high-order g modes, which probe the deep stellar interior and are highly sensitive to near-core physics \citep[e.g.,][]{2015ApJS..218...27V,2019A&A...626A.121O}. Although asymptotic theory predicts nearly uniform period-spacing patterns, these structures are inherently modified by rotation, chemical gradients, and internal mixing, providing powerful diagnostics of angular-momentum transport and mixing processes \citep{2015A&A...574A..17V, 2016A&A...593A.120V, 2021NatAs...5..715P}. This field was revolutionized by high-precision, spaceborne photometry, initiated by \textit{Kepler} \citep{2010Sci...327..977B} and subsequently expanded by the \textit{Transiting Exoplanet Survey Satellite} \citep[TESS;][]{2015JATIS...1a4003R}. These dense and complex Fourier spectra enable systematic ensemble studies, successfully deriving buoyancy properties and near-core rotation rates for hundreds of $\gamma\text{~Dor}$ stars \citep{2018A&A...618A..47C, 2020MNRAS.491.3586L, 2022A&A...668A.137G}.

However, the accuracy of these asteroseismic constraints hinges on identifying the intrinsic patterns of $g$ modes, a process potentially vulnerable to frequency contamination. Previous studies have systematically examined the external contamination of super-Nyquist frequency aliases on these $g$-mode patterns \citep{2025A&A...693A..63W}, confirming that their impact is negligible in $\gamma$ Dor stars. Nevertheless, given the extraordinary richness of $\gamma$ Dor frequencies, they inevitably present a number of linear combination frequencies, a phenomenon widely observed across various classes of pulsating stars \citep{2015MNRAS.450.3015K,2020MNRAS.498.1194L}. These combinations can complicate intrinsic $g$-mode patterns, yet their exact impact remains unexplored. Furthermore, $\gamma$ Dor stars serve as an ideal testbed for investigating the real origins of these combinations, which can provide critical insights into those observed in all other types of pulsators.

Linear combination frequencies ($f_{\text{comb}} = \sum n_i f_i$, including harmonics and sum/differences) are dependent peaks in amplitude spectra \citep{1999A&A...349..225B}. In this work, we adopt the definition and identification criteria described in Appendix~\ref{app:method}. They must be identified and removed to prevent misidentification before seismic analysis and modeling \citep{2020MNRAS.498.1194L}. Previous work attributes the origins of these frequencies to two main mechanisms: artifacts induced by the non-linear distortion of light curves, and an increase in detectable amplitude of intrinsic modes under non-linear resonance conditions. In the former regime, linear combinations were proposed by earlier studies to arise from scenarios such as the response to surface convection or local surface temperature during pulsation cycles \citep[see, e.g.,][]{1992MNRAS.259..519B,1995ApJS...96..545B,2001MNRAS.323..248W}. As predicted by the analytical formula, these artifacts can serve as mode discriminants by inherently conveying the information of their parent modes in pulsating white dwarfs \citep{2001MNRAS.323..248W,2005ApJ...635.1239Y}, yet they fail in pulsating main-sequence stars, due to the relatively long response timescales \citep{2012MNRAS.422.1092B}. Nevertheless, harmonic frequencies have recently been proposed to be helpful for 
constraining internal convective dynamics in RR~Lyrae stars \citep{2026A&A...710L..23N}. Within the latter regime, should the frequency of even a small-amplitude pulsation mode coincide with the resonance conditions of two or three other modes via linear combinations, the subsequent non-linear energy exchanges will drive these hidden modes to observable amplitudes \citep[see, e.g.,][]{1985AcA....35..229M,1997A&A...321..159B}. These expectations have been intensively investigated from both observational and theoretical perspectives to characterize the amplitude and frequency modulations, as well as the amplitude saturations, of resonant coupling modes. This ongoing research spans various classes of pulsators, from $\delta$~Scuti \citep{2014ApJ...783...89B,2023ApJ...950....6M} and slowly pulsating B stars \citep[SPB;][]{2024A&A...687A.265V} to hot B subdwarfs \citep[sdB;][]{2016A&A...594A..46Z} and white dwarfs \citep{2016A&A...585A..22Z}.

In this work, we conducted a comprehensive survey of linear-combination frequencies in 608 $\gamma$~Dor stars observed by Kepler \citep{2020MNRAS.491.3586L}. We aim to quantify the prevalence of these dependent frequencies and evaluate their contribution to the observed amplitude spectra. 
We present the main results in Section~\ref{sec:results}, reporting the statistical properties of the linear combinations. Finally, Section~\ref{sec:discussion} discusses these findings and summarizes the main conclusions of this work.

\section{Statistical results}
\label{sec:results}

\begin{figure}[!htb]
    \centering
    \includegraphics[width=\columnwidth]{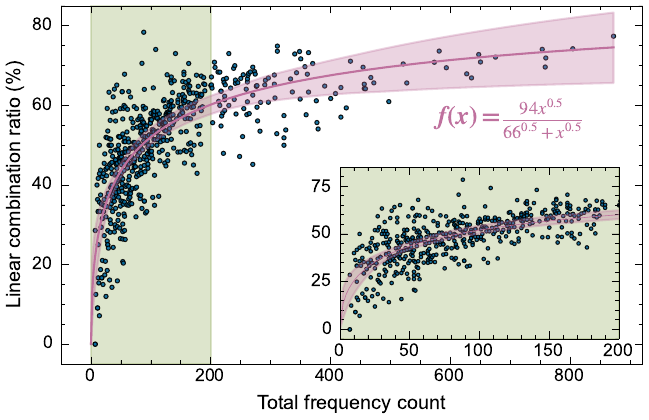}
   \caption{Rate of linear combination frequencies as a function of the total number of detected frequencies in each $\gamma$~Dor star. The solid curves are fitted with the Hill function, with the $3\sigma$-uncertainty limits indicated by the shaded area. The inset panel highlights the region representing stars with fewer than 200 detected frequencies.}
    \label{fig:linearcom_ratio}
\end{figure}

\begin{figure}[htb]
    \centering
    \includegraphics[width=\columnwidth]{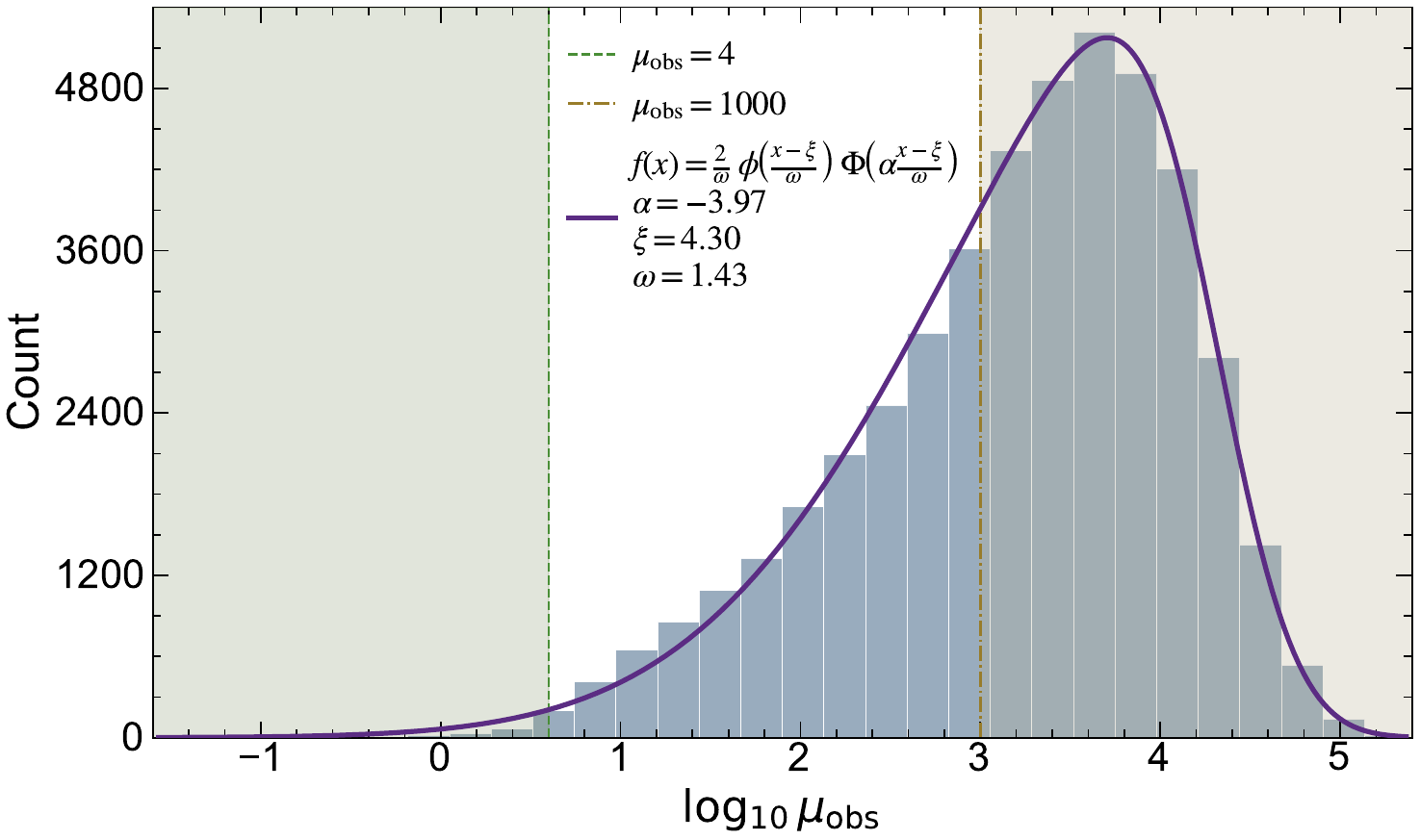}
\caption{Distribution of the amplitude ratio $\mu$ for candidate combination frequencies in the $\gamma$~Doradus sample. The green dashed and ochre dash-dot lines indicate $\mu=4$ and $\mu=1000$ (see text for details), respectively. The solid curve shows the left-skewed fit, with the parameters given in the figure.}
    \label{fig:amplitude_ratio}
\end{figure}

\begin{figure}[!htb]
    \centering
    \includegraphics[width=\columnwidth]{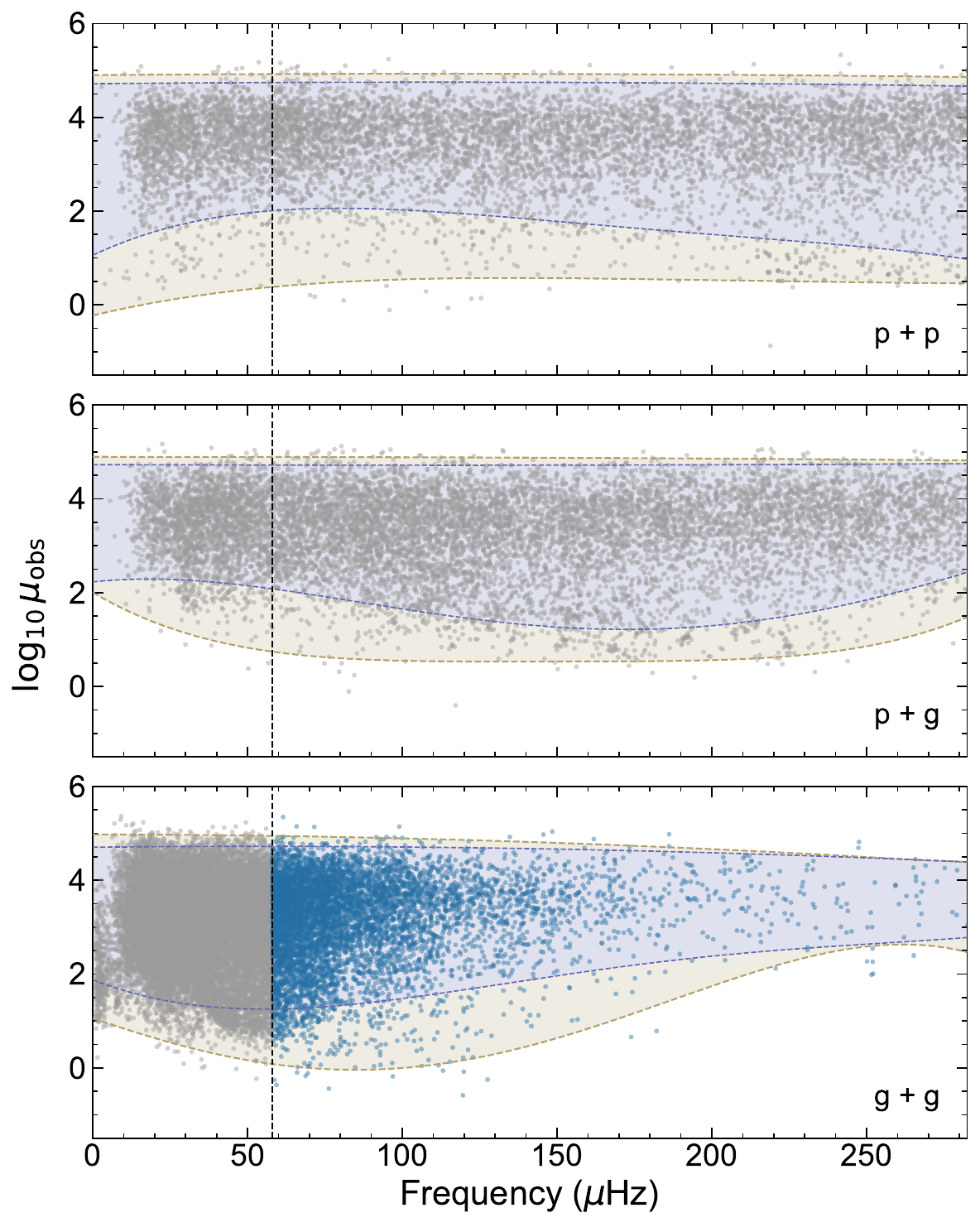}
    \caption{Distribution of the observed amplitude ratio $\mu_{\rm obs}$ for linear combination frequencies as a function of frequency.  From top to bottom, the three panels separate the candidates according to the nature of their parent modes: two $p$-modes, one $p$-mode and one $g$-mode, and two $g$-modes, respectively. The vertical dashed line indicates the adopted boundary between the $g$-mode and $p$-mode region at 58~$\mu$Hz. 
    The shaded regions represent the smoothed empirical envelopes of the distribution: the outer (yellow) and inner (purple) regions correspond to the [0.1\%, 99.8\%] and [5\%, 99\%] intervals, respectively.
   }
    \label{fig:pgmode}
\end{figure}

After photometry processing, we applied the procedures for frequency extraction and the identification of linear combination frequencies, as described in Appendix~\ref{app:method}, to the full sample of 608 $\gamma$~Doradus stars. In total, 47,863 frequencies meet our criteria for linear combination frequencies out of the entire catalog of 81,265 frequencies, including 1,895 harmonics and 45,968 combinations involving two parent frequencies. For most $\gamma$ Dor stars, we typically detect between 30 and 300 significant frequencies per target.

Figure~\ref{fig:linearcom_ratio} shows the occurrence rate of linear combination frequencies for each $\gamma$~Dor star. A distinct correlation can be seen as the combination fraction exhibits a gradual increase with the growing number of detected frequencies. This rising trend reaches a plateau around 200 frequencies where the fraction begins to level off at around $75\%$. We fit the distribution using a Hill function, which is expressed as $f(x) = ax^n / (b^n + x^n)$, where $x$ represents the total number of frequency count. The best-fit coefficients $a$, $b$ and $n$ are provided in the figure.

To quantify the relative amplitudes of the linear combination frequencies, we calculate the amplitude ratio for each of them following \citet{2014ApJ...783...89B}:
\begin{equation}
\mu_{\rm obs} = \frac{A_c}{A_a A_b},
\end{equation}
where $A_c$ is the fractional amplitude of the combination frequency, and $A_a$ and $A_{b}$ are those of the two parent frequencies.

Figure~\ref{fig:amplitude_ratio} shows the distribution of $\mu_{\rm obs}$ for the identified combination frequencies in a logarithmic scale. 
The distribution is characterized by a simple left-skewed shape, which agrees well with our best-fit results with parameter provided in the figure. This profile exhibits a clear peak at $\log_{10}\mu_{\rm obs} \simeq 3.63$ (i.e., $\mu_{\rm obs} \approx 4.3 \times 10^{3}$) and a median value of $\log_{10}\mu_{\rm obs} \simeq 3.34$ (i.e., $\mu_{\rm obs} \approx 2.2 \times 10^{3}$), indicating that the bulk of the population clusters within this regime. Specifically, most linear combinations lie in the range of $\mu_{\rm obs} \sim 10^2\text{--}10^4$, which includes 31,907 cases and accounts for 69.41\% of the entire sample. At the lower end, we observe only 177 combinations with $\mu_{\rm obs} < 4$, which represents a mere 0.4\% of the sample. In contrast, at the higher end, a substantial population of 29,411 combinations extends beyond $\mu_{\rm obs} \gtrsim 10^{3}$, accounting for 64\%. These two boundaries, $\mu_{\rm obs} \simeq 4$ and $\mu_{\rm obs} \simeq 10^3$, correspond to the thresholds defined by \citet{2014ApJ...783...89B} for linear combinations induced by nonlinearity flux distortion and nonlinear resonant couplings, respectively.

Figure~\ref{fig:pgmode} compares the observed amplitude ratio, $\mu_{\rm obs}$, of linear combination frequencies with different types of parent-modes. For simplicity, we merely divided the frequencies of p- or g-modes by 58~$\mu$Hz \citep{2011A&A...534A.125U}. The observed $\mu_{\rm obs}$ exhibits a distinct upper bound $\sim10^5$ across all three types of parent modes. At the lower bound, the linear combinations with $p+p$ and $p+g$ parents behave similarly. However, a slight difference emerges below $10^2$, where the former exhibits a sparse distribution, while the latter still maintains a small fraction of cases.
In contrast, a notable difference appears in combinations with $g+g$ parents. We observe that the lower bound of $\mu_{\rm obs}$ is systematically elevated with increasing frequency above $80~\mu$Hz, although the population is limited. Moreover, at lower frequencies $<80~\mu$Hz, the lower bound of $\mu_{\rm obs}$ is relatively lower across all types of these distributions.

\section{Discussion}
\label{sec:discussion}
We have completed the first systematic survey to estimate the rate of significant signals in Fourier spectra being linear combination frequencies in $\gamma$ Dor stars from Kepler photometry. As shown in Fig.\,\ref{fig:linearcom_ratio}, the rate increases with the detected frequency number per star before leveling off at an upper ceiling $\sim75\%$. This trend is mathematically expected, as a larger pool of detected frequencies within a fixed frequency range (up to the Nyquist frequency of $\sim282.5~\mu$Hz) inherently increases the probability of identifying coincidental linear combinations within the uncertainty of the measurement, leading to the observed saturation fitted by the Hill function. This significant portion of frequencies in linear combinations indicates that caution must be exercised during mode identification prior to subsequent seismic analysis. Although chance coincidences are mathematically favored, the precise alignment of these signals within the remaining frequency space points to a physical origin that cannot be uncovered by frequency relations alone, necessitating a physical interpretation based on additional physical constraints, such as the amplitude ratio and its distribution pattern.

One key factor is that the observed amplitude ratio, $\mu_{\rm obs}$, shows a broad distribution, mostly concentrated around $100$–$10,000$ (see Fig.\,\ref{fig:amplitude_ratio}). These values are significantly higher than those predicted by nonlinearity surface distortion in temperature \citep{1995ApJS...96..545B} or convective response \citep{2001MNRAS.323..248W}. In their calculations, $\mu_{\rm obs}$ is expected to be on the order of 10 or less for most cases, even after taking geometric effects into account. This implies that these conventional nonlinearity distortions can only account for a small portion of the detected linear combinations in our $\gamma$~Dor sample. In addition, \citet{2012MNRAS.422.1092B} noted that the behavior of nonlinear flux distortion is somewhat different when extending the framework from white dwarfs to main-sequence pulsators like $\gamma$~Dor stars.

However, these large $\mu_{\rm obs}$ values can be accommodated naturally if we consider nonlinear resonance as the potential mechanism \citep{1982AcA....32..147D}. Recently, significant attempts have been made at seismic diagnostics of SPB stars via amplitude ratios derived from three-g-mode resonance, moving beyond conventional linear frequency constraints \citep{2024A&A...687A.265V}. Similarly, frameworks of nonlinear resonance have been applied to interpret $\mu_{\rm obs}\sim10^3-10^4$ in a specific $\delta$ Scuti star KIC~8054146 \citep{2014ApJ...783...89B,2023ApJ...950....6M}. Specifically, the amplitude ratios for three-mode direct resonance calculated by \citet{2023ApJ...950....6M}, which span from $\sim 10^2$ to nearly $10^6$, agree well with our $\mu_{\rm obs}\sim 10^2-10^4$ most observed in $\gamma$ Dor stars, although their model was specifically developed for $\delta$~Scuti stars. For a simple comparison, taking one of their model examples with parameters $M=1.85\,M_\odot$, $T_{\rm eff}=7350\text{ K}$, and $\log g=3.96$, the amplitude ratios were derived spanning $\log_{10}\mu\simeq3.5\text{--}5.9$, a range that falls well within our observed interval of $\log_{10}\mu\simeq-1\text{--}5.2$. The similarity in parameters of some $\delta$ Scuti and $\gamma$ Doradus stars implies that nonlinear resonance can qualitatively account for our observed amplitude ratios. However, a more rigorous explanation warrants dedicated stellar models of $\gamma$ Dor to solve the nonlinear amplitude equations in future studies. 

Assuming nonlinear resonance is a plausible explanation for our results, our observed amplitude ratio distributions, statistically, for resonances between different mode types (Fig.\,\ref{fig:pgmode}) could provide potential physical implications. 
 Within the framework of direct resonance, Eqs.\,(\ref{eq:22}) and (\ref{eq:33}) show that $\kappa_{abc}$ strongly depends on the integral of the kernel function involving the parent and child modes. This dependence is critical for understanding the distribution characteristics shown in Fig.\,\ref{fig:pgmode} and the frequency detuning in Fig.\,\ref{fig:detuning}, particularly the deficit in high-frequency combinations involving two g-mode parents. Since p- and g-modes are sensitive to different regions of the stellar interior, their spatial overlap is limited; consequently, detecting these modes through nonlinear coupling of $ggp$ or $ppg$ becomes relatively difficult. Note that the number of $ggp$ combinations is much smaller than that of $ggg$. Moreover, $ggp$ couplings are significantly weaker, likely because their kernel integral is much smaller compared to other types of coupling. Such interactions may involve not only second-order perturbations but also higher-order ones, such as third-order combinations that mediate multi-mode resonant interactions \citep{1995A&A...296..405B}. In addition, we tentatively identify potential parametric resonance candidates based on their amplitude ratios, guided by the numeric explorations discussed in \citet{2025ApJ...986...33M} and \citet{2024A&A...687A.265V}. However, given the simplicity of our amplitude-ratio approach, only a small fraction of the combinations satisfy these criteria at this stage (see details in Appendix~\ref{app:parametric}). While a more detailed physical investigation is beyond the scope of this study, 
our findings provide a promising space for further exploration and advancement in nonlinear resonance.

Now, both physical and mathematical evidence suggests that most linear combination frequencies in $\gamma$ Dor stars can be attributed to nonlinear resonant modes, which means that these child modes exist as physical entities. Mathematically, taking frequency uncertainties into account, one would expect only a few accidental linear combinations within a sample of ~200 frequencies, far below the 60\% observed in $\gamma$ Dor stars. In fact, they outnumber the mere $\ell=1$ or $\ell=2$ eigenmodes available in this small frequency range. 
Alternatively, they may be associated with higher-degree eigenmodes whose amplitudes are enhanced by resonant condition subject to selection rules. This possibility remains speculative, although Kepler provides such tentative indications in sdB stars \citep[e.g.,][]{2013AcA....63...79B}.
To extend this work, the recently discovered wildly oscillating AF stars, which exhibit complex ridge-like pulsation spectra \citep{2026A&A...712A.220A}, could serve as ideal candidates for further investigation.

Bearing in mind that for any further investigation of linear combinations, the noise floor will inevitably alter the detection rate of combination frequencies, with this rate decreasing systematically as the signal-to-noise ratio (S/N) drops (see Fig.\,\ref{fig:snr}). To mitigate this effect, future studies should aim for maximum completeness in frequency extraction, as missing low-amplitude frequencies could affect the final conclusions. Beyond these observational limits, our framework currently implies that child frequencies possess lower amplitudes, a trend visually confirmed by the amplitude distribution in Fig.\,\ref{fig:amp_fre}. Nevertheless, nonlinear resonance allows a child mode to exceed its parents in amplitude. Accounting for this possibility will inevitably alter the true distribution of the amplitude ratio, thereby bringing our results into closer alignment with those reported by \citet{2023ApJ...950....6M}. Crucially, the accuracy of mode identification inevitably affects the robustness of inferred seismic properties and subsequent modeling, such as period spacings and rapid rotation splittings.

We conclude this letter by anticipating a bright future for the study of nonlinear resonance in diverse types of pulsating stars across the Hertzsprung-Russell diagram. While these statistical results show a good starting point, detailed models in the future will offer more precise solutions for nonlinear dynamics, potentially providing a way to understand amplitudes within the nonlinear regime. Undeniably, as another critical expectation of nonlinear resonance, amplitude and frequency modulations are essential in this field as well, which will be characterized across a larger sample of pulsators and by future and ongoing space missions.

\begin{acknowledgements}
The authors acknowledge the support from the National
Natural Science Foundation of China (NSFC) through grant Nos. 12273002,
12541303, 12673036 and 12427804, the Beijing Nova Program No. 202604841350, the Central Guidance for Local Science and Technology Development Fund under No. ZYYD2025QY27 and the International Centre of Supernovae (ICESUN), the Yunnan Key Laboratory of Supernova Research (No. 202505AV340004). This work is partially supported by the science research grants from the China Manned Space Project No.~CMS-CSST-2025-A13. J.N.F. acknowledges support from the Tianchi Talent Introduction Plan. S.C. acknowledges support from the Centre National d’Etudes Spatiales (CNES, France), focused on the Kepler Mission. All of the Kepler data used in this paper can be found in MAST. The authors appreciate all who have contributed to making these missions possible. Funding for the Kepler Mission is provided by NASA’s Science Mission Directorate. 
\end{acknowledgements}

\bibliographystyle{aa} 
\bibliography{ref}

\FloatBarrier 
\clearpage

\begin{appendix}

\nolinenumbers

\section{Data and methods}
\label{app:method}

\begin{figure*}
    \sidecaption
    \includegraphics[width=12cm]{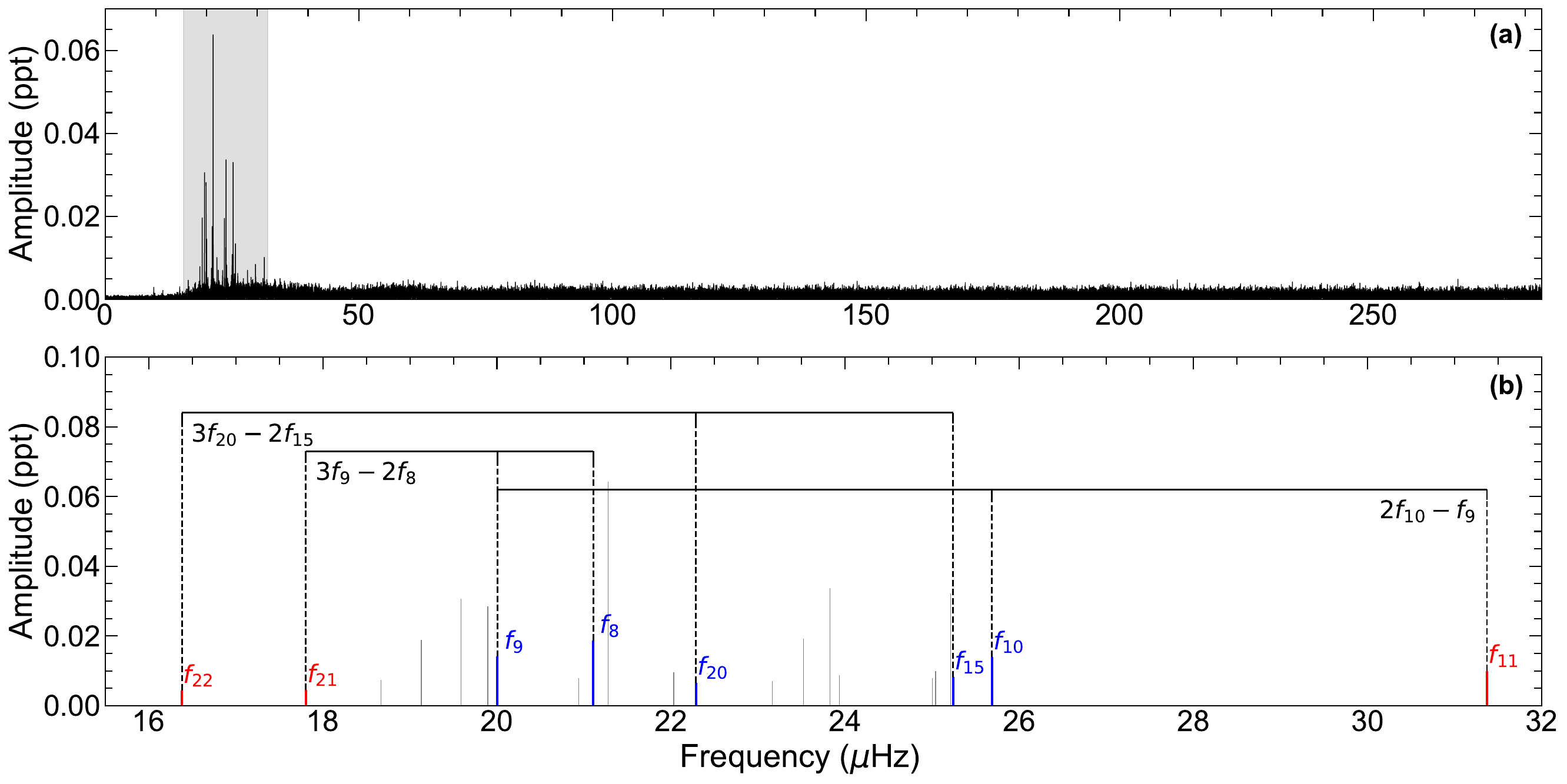}
    \caption{Frequency analysis and linear combination identification for the representative $\gamma$~Dor star KIC~3238245.
(a) The entire Lomb-Scargle periodogram, with the gray shaded region marking the frequency interval enlarged in panel (b). (b) Zoomed-in view of the selected frequency interval. Blue, red, and gray vertical lines mark the parent frequencies, linear combination frequencies, and other detected signals within the same interval, respectively. The labels and horizontal segments indicate the frequency IDs and the corresponding linear combination relations.}
    \label{fig:sample}
\end{figure*}

The photometric data analyzed in this work are drawn from \citet{2025A&A...693A..63W}, who compiled a sample of 608 $\gamma$~Dor stars from the catalog of \citet{2020MNRAS.491.3586L}. Details concerning the light curve assembly and reduction procedures can be found therein. Unlike their approach, which maintained a stringent S/N threshold of 8.0 for their specific scientific aims, we adopt a lower threshold of $\text{S/N} \ge 5.0$ during frequency extraction to ensure a more comprehensive sample using the \texttt{FELIX} code (see details in \citeauthor{2010A&A...516L...6C} \citeyear{2010A&A...516L...6C} and \citeauthor{2016A&A...585A..22Z} \citeyear{2016A&A...585A..22Z}). Following our previous work, we used the \texttt{sLSP4SNFs} pipeline to remove super-Nyquist frequency ambiguities \citep{2026A&A...710A.245M}. Consequently, we obtained a final sample of 81,265 significant frequencies in 608 $\gamma$~Dor stars for our subsequent analysis.

The same iterative framework was then applied to identify harmonic and combination frequencies in the final frequency catalog. For each star, the frequencies were sorted in descending order of amplitude, with higher-amplitude frequencies preferentially considered as candidate parent frequencies, while lower-amplitude frequencies were tested as potential harmonics or linear combinations. 
The identification procedure first searched for harmonic relations of the form $f_k \approx m f_i$, where $m$ is an integer satisfying $2 \le m \le 10$. A candidate frequency was accepted as a harmonic if 
\begin{equation}
|f_k - m f_i| \le 3 \sqrt{\sigma_k^2 + (m \sigma_i)^2},
\end{equation}
where $\sigma_i$ and $\sigma_k$ represent the uncertainties of the frequencies $f_i$ and $f_k$, respectively.
We then searched for two-parent combination frequencies of the form $f_k \approx m f_i + n f_j$, where $m,n = \pm1, \pm2, \pm3$, retaining only positive resulting frequencies. A candidate was accepted if
\begin{equation}
|f_k - (m f_i + n f_j)| \le
3 \sqrt{
\sigma_k^2 +
(m \sigma_i)^2 +
(n \sigma_j)^2
}.
\end{equation}

Only frequencies with amplitudes higher than that of the target frequency were considered as potential parent candidates. When multiple valid solutions were found, the preferred identification was selected based first on the lowest total combination order ($|m| + |n|$), followed by the smallest normalized frequency deviation.

Figure~\ref{fig:sample} presents a representative pulsation spectrum of the $\gamma$~Dor star KIC~3238245, illustrating the identified parent frequencies and their associated linear combination components.

\section{Distribution of the frequency detuning}
\label{detuning}

\begin{figure}[!htb]
    \centering
    \includegraphics[width=\columnwidth]{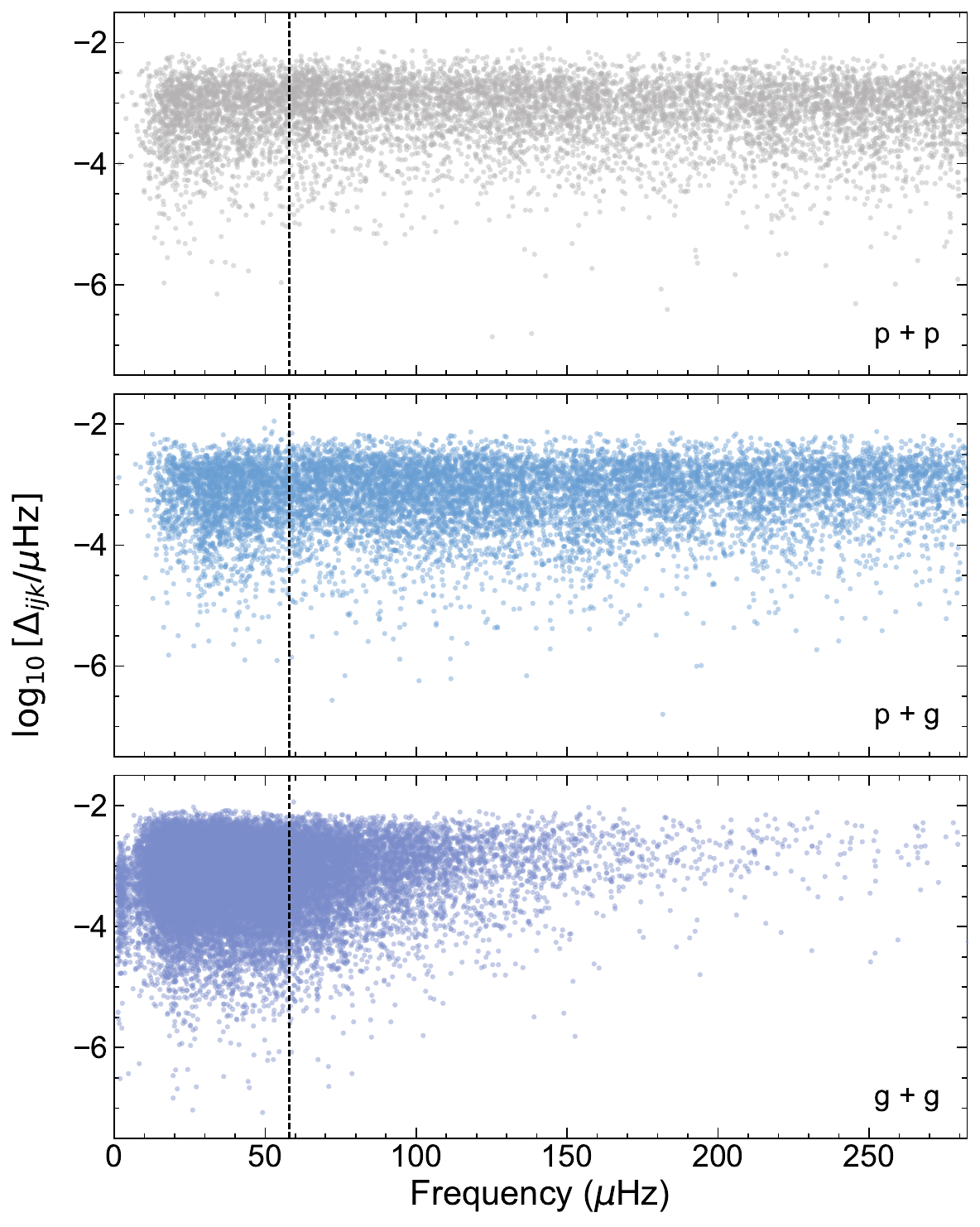}
    \caption{Distribution of the frequency detuning $\Delta_{ijk}$ as a function of frequency for the identified linear combination frequencies. The panels follow the same $p+p$, $p+g$, and $g+g$ parent-mode
classification as in Fig.~\ref{fig:pgmode}. The vertical dashed line marks the adopted boundary between the
$g$-mode and $p$-mode regions at 58~$\mu$Hz.} 
    \label{fig:detuning}
\end{figure}

To further examine how closely the identified combination frequencies satisfy their corresponding resonance conditions, we calculate the frequency detuning for each identified combination frequency as
\begin{equation}
\Delta_{ijk}
=
\left|f_k-(m f_i+n f_j)\right|.
\end{equation}
Figure~\ref{fig:detuning} shows that the frequency detunings of the identified combination frequencies are strongly concentrated toward small values.
The detuning remains below
$0.012~\mu$Hz for all identified combinations. 
The overall detuning ranges are similar, although the identified combination frequencies with $g+g$ parents show a different distribution profile.

\section{Amplitude ratio formula of direct resonance}
\label{app:direct}
In nonlinear coupling formalism (e.g., \citealt{2023ApJ...950....6M}), the amplitude ratio between three mode interaction can be written as,
\begin{equation} \label{eq:22}
    \mu =
    \frac{|\omega_c \kappa_{abc}|}
    {\sqrt{\Delta_{abc}^2+\gamma_c^2}},
\end{equation}
where $\omega_c$ and $\gamma_c$ are the angular frequency and the damping rate of the child mode, respectively, and the frequency detuning $\Delta_{abc} = |\omega_a \pm \omega_b -  \omega_c|$. The nonlinear coupling coefficient $\kappa_{abc}$ is calculated through
\begin{equation} \label{eq:33}
    \kappa_{abc} =\frac{1}{E_\star}
\int d^3x\,
\boldsymbol{\xi}_a
\cdot
\boldsymbol{f}_2\!\left[\boldsymbol{\xi}_b,\boldsymbol{\xi}_c\right],
\end{equation}
where $\boldsymbol\xi_{a,b,c}$ are the linear eigenfunctions of the three modes involved, $\boldsymbol{f}_2$ is the second-order operator representing the nonlinear term, and $E_\star$ is the characteristic energy of the star.

\section{Amplitude ratios for identifying parametric resonance candidates} \label{app:parametric}

\begin{figure}[!htb]
    \centering
    \includegraphics[width=\columnwidth]{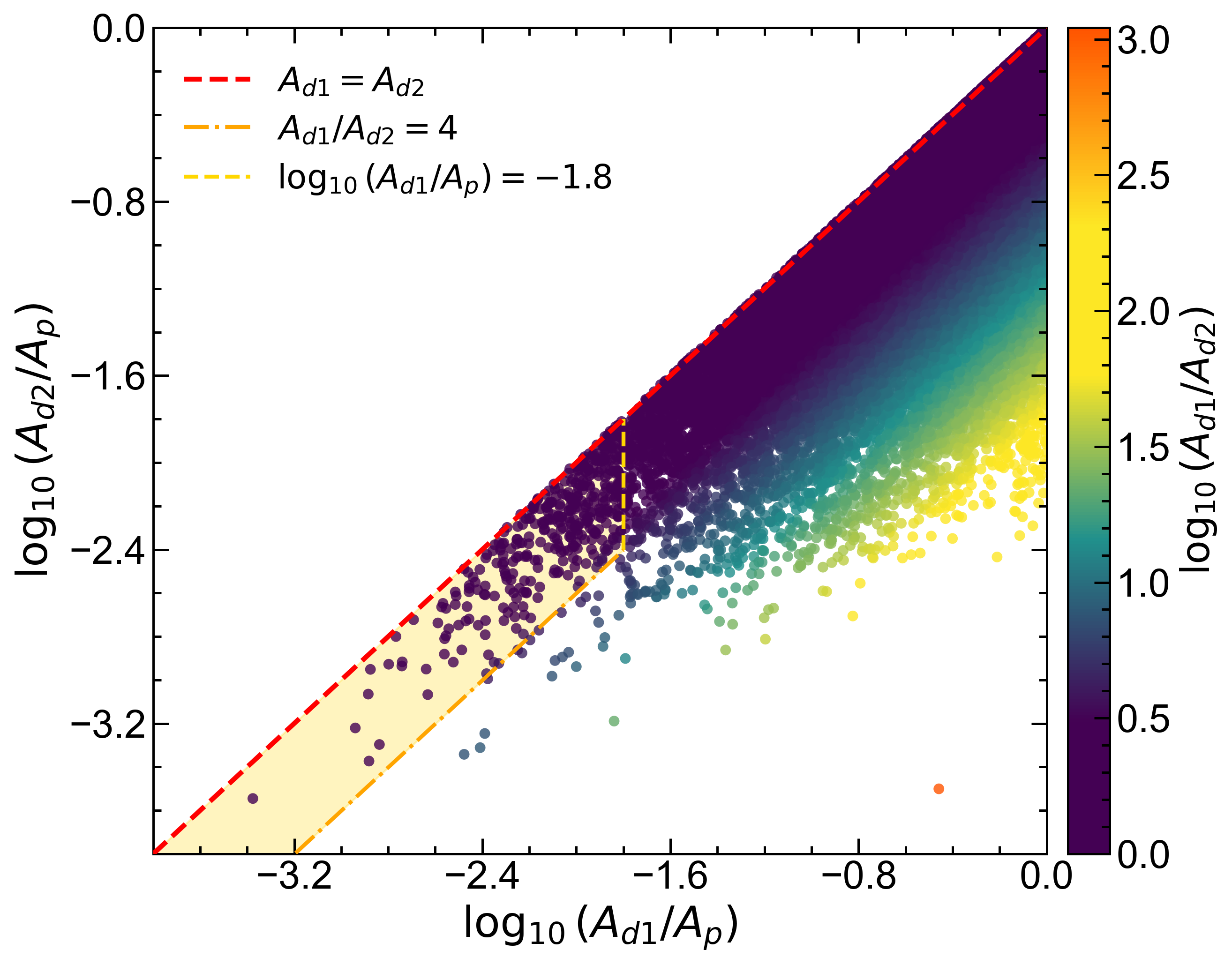}
    \caption{
    Observed daughter-to-parent amplitude ratios $A_d/A_p$ as an indicator to identifying candidate of parametric resonances. Each point represents a combination frequency, with its position determined by the two daughter-to-parent amplitude ratios. The color indicates the daughter-to-daughter amplitude ratio, while the red dashed line denotes the bisector corresponding to equal amplitudes of the two daughter modes. The orange dash-dotted line denotes $A_{d1}/A_{d2}=4$, while the golden dashed line denotes $\log_{10}(A_{d1}/A_p)=-1.8$. The light-yellow region highlights potential parametric resonance candidates. These threshold values are chosen somewhat arbitrarily, with general guidance from \citet{2024A&A...687A.265V} and \citet{2025ApJ...986...33M}, rather than being derived from a specific physical criterion.}
    \label{fig:para}
\end{figure}

To investigate parametric resonances in our results, we considered only those combination frequencies that satisfy
\begin{equation}
    f_k = m f_i \pm n f_j,
\end{equation}
where $m,n=\pm1$. Following the procedures described in Appendix~\ref{app:method}, we identified 27,231 combination frequencies, with the highest-amplitude mode designated as the parent and the other two modes as the daughter modes. We calculate the daughter-to-parent amplitude ratios, $A_{d1}/A_p$ and $A_{d2}/A_p$, where $A_{d1}\geq A_{d2}$, as well as the daughter-to-daughter amplitude ratio, $A_{d1}/A_{d2}$. 

The results of this analysis are presented in Fig.~\ref{fig:para}, which shows the distribution of the combination frequencies in the plane defined by the two daughter-to-parent amplitude ratios. The two daughter modes become increasingly similar in amplitude as $A_{d1}/A_{d2}$ approaches unity, with points lying closer to the red dashed bisector. According to the description in \citet{2024A&A...687A.265V}, parametric resonance is expected to involve a high-amplitude parent mode that is resonantly coupled two daughter modes with lower amplitudes. Based on the amplitude ratios between the parent and daughter modes, we tentatively identify 415 out of the 27,231 combinations (1.52\%) as potential parametric resonance candidates. However, confirming the physical nature of candidates would require more detailed modeling and calculating the coupling parameters. Such a detailed physical analysis is beyond the scope of the present statistical study.

\section{Rates of linear combinations with S/N}
\label{app:snr}

\begin{figure}[!htb]
    \centering
    \includegraphics[width=\columnwidth]{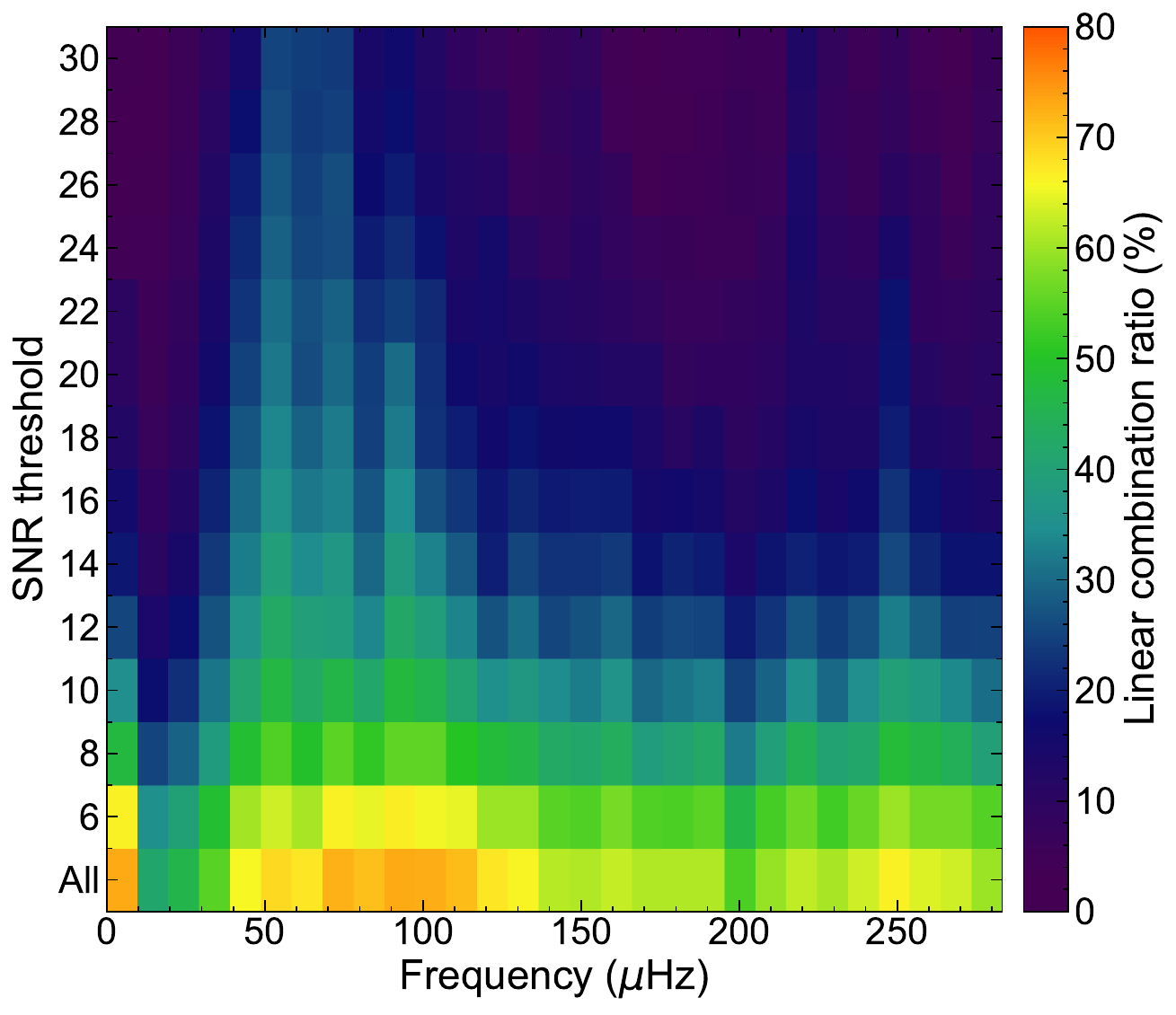}
    \caption{Rate of linear combination frequencies as a function of frequency and S/N threshold. Here, linear combinations include both harmonics and combinations of two independent frequencies. The color scale gives the percentage of detected frequencies identified as linear combinations in each frequency bin. The bottom row labeled ``All'' represents the full frequency sample, while the other corresponds the rate after applying progressively higher S/N thresholds.}
    \label{fig:snr}
\end{figure}

Figure~\ref{fig:snr} shows the fraction of linear-combination
frequencies varying with frequency and S/N threshold. In the full sample, such candidates represent a substantial fraction of the detected frequencies over a broad frequency range. As progressively higher S/N thresholds are applied, the fraction generally decreases across most frequency bins, although the detailed pattern is not uniform over frequency. This suggests that the inferred contribution of linear-combination frequencies depends on the adopted S/N threshold, with lower S/N samples containing a larger relative fraction of such candidates. Nevertheless, linear-combination candidates remain present over a wide frequency range even after applying stricter S/N cuts, indicating that their occurrence is not limited to a narrow frequency interval.

\section{Distribution of amplitude and frequency}
\label{app:ampfreq}

\begin{figure}[!htb]
    \centering
    \includegraphics[width=\columnwidth]{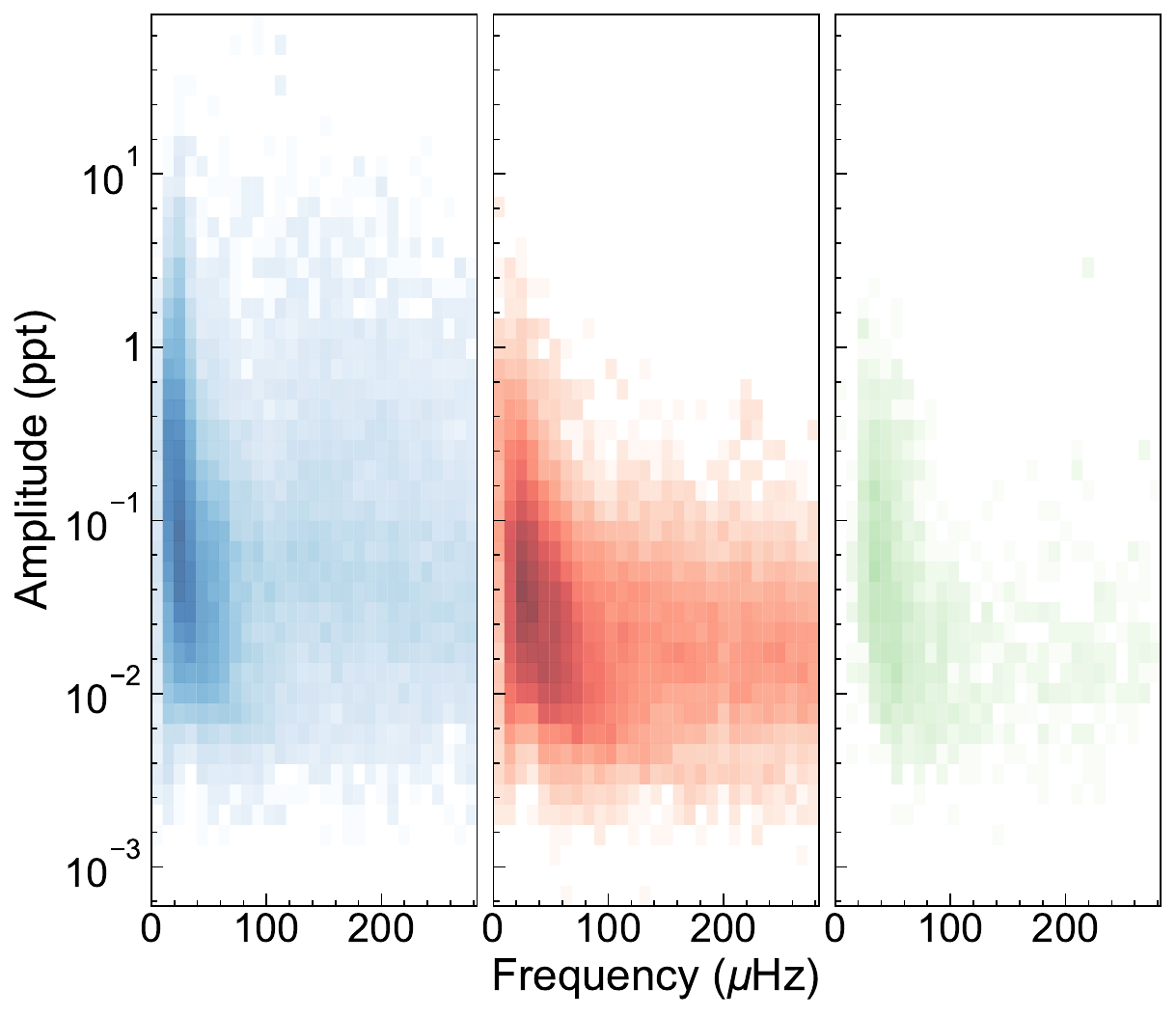}
    \caption{Distribution of amplitudes and frequencies for all detected signals across the 608~$\gamma$~Dor sample.  From left to right, the panels represent intrinsic pulsation, linear combination frequencies, and harmonics, respectively. The color intensity denotes the signal density within each frequency–amplitude bin, with darker shades corresponding to higher counts. }
    \label{fig:amp_fre}
\end{figure}

Figure~\ref{fig:amp_fre} shows amplitude and frequency of linear combination frequencies extend over a broad frequency interval but are preferentially found at lower amplitudes. Harmonics constitute a much smaller population and are concentrated mainly at low amplitudes and low frequencies, whereas frequencies not identified as combinations or harmonics span the widest range and dominate the low-frequency g mode region.

\end{appendix}

\end{document}